\documentclass[seceq]{ptptex}

\usepackage{graphicx}
\usepackage{wrapft}
\usepackage{amsmath,amsfonts,bm,amssymb}

\newcommand{\be}{\begin{equation}}
\newcommand{\ee}{\end{equation}}
\newcommand{\bea}{\begin{eqnarray}}
\newcommand{\eea}{\end{eqnarray}}

\newcommand{\eq}[1]{Eq.~(\ref{eq:#1})}
\newcommand{\sect}[1]{Sec.~\ref{sec:#1}}

\newcommand{\fig}[1]{Fig.~\ref{fig:#1}}
\newcommand{\tabl}[1]{Table~\ref{table:#1}}

\newcommand{\del}{\partial}
\newcommand{\tpi}{\tau_{\pi}}
\newcommand{\tPi}{\tau_{\Pi}}
\newcommand{\tR}{\tau_J}
\newcommand{\fm}{{\rm fm}}

\bmdefine{\bmp}{ \bm{p} }
\bmdefine{\bms}{ \bm{s} }
\bmdefine{\bmT}{ \bm{T} }
\bmdefine{\bmeps}{ \bm{\epsilon} }

\title{
String theory implications on causal hydrodynamics%
}

\author{
Makoto \textsc{Natsuume}\footnote{E-mail: makoto.natsuume@kek.jp}%
}

\inst{
Theory Division, Institute of Particle and Nuclear Studies, \\
KEK, High Energy Accelerator Research Organization,\\ 
Tsukuba, Ibaraki, 305-0801, Japan
}

\abst{
We summarize the main lessons for causal hydrodynamics/second order hydrodynamics/Israel-Stewart theory from string theory. 
Based on an invited talk presented at NFQCD2008 Symposium (March 3-6, 2008, YITP). 
}

\begin{document}

\maketitle

\section{String theory and quark-gluon plasma}

Causal hydrodynamics is one of the most active areas in the studies of quark-gluon plasma (QGP), and I will tell you the implications of string theory on this issue \cite{Natsuume:2007ty,Natsuume:2007tz,Natsuume:2008iy}.

I first review the relation between QGP and string theory quickly in this section. Then, I explain the basic idea of causal hydrodynamics pedagogically in \sect{causal}. Our aim is to study causal hydrodynamics from string theory which is described in \sect{ads_causal}. Actually, a number of papers appeared recently on this issue\cite{Benincasa:2007tp,Baier:2007ix,Bhattacharyya:2007jc}.%
\footnote{See also Refs.~\citen{Loganayagam:2008is,VanRaamsdonk:2008fp,Bhattacharyya:2008xc,Amado:2008ji,Bhattacharyya:2008ji,Buchel:2008bz,Kapusta:2008ng,Haack:2008cp}.} 
In particular, our paper has many overlaps with Ref.~ \citen{Baier:2007ix}.

Our main tool is the so-called AdS/CFT duality \cite{Aharony:1999ti}. The AdS/CFT duality claims that a finite temperature gauge theory at strong coupling is dual to a black hole. A black hole appears since a black hole is a thermal system as well. Due to the Hawking radiation, a black hole has the notion of temperature: this is the reason why there can be a correspondence between these two in the first place. Since QGP is exactly a finite temperature gauge theory system at strong coupling, our aim is to study QGP using black holes. 

\begin{wrapfigure}{r}{6.6cm}
\begin{center}
\includegraphics{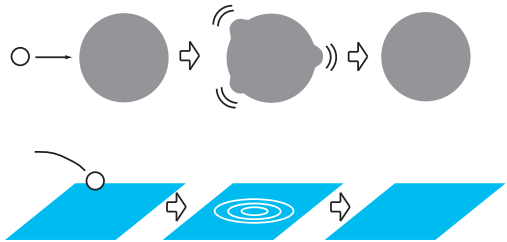}
\caption{When one adds a perturbation to a black hole, the black hole behavior is similar to a hydrodynamic system. In hydrodynamics, this is a consequence of  viscosity.}
\label{fig:hydro}
\end{center}
\end{wrapfigure}

According to RHIC experiment, QGP behaves like a fluid with a very low viscosity. Then, the AdS/CFT duality implies that a black hole also behaves like a fluid with a low viscosity. In fact, black holes and hydrodynamic systems behave similarly (\fig{hydro}). For example, consider a water pond and drop some object. Then, you generate surface waves, but they decay quickly, and the water pond returns to a state of stable equilibrium. This is a dissipation phenomenon; in hydrodynamics, the dissipation is a consequence of viscosity. 

Black holes behave similarly. Again drop some object to a black hole. Then, the shape of the black hole horizon is distorted, but such a perturbation is not stable. It decays quickly, and the black hole returns to the original symmetric shape. If you regard this as a dissipation as well, the dissipation occurs in this case since the perturbation is absorbed by the black hole. Thus, you can consider the notion of viscosity for black holes as well, and the ``viscosity" for black holes should be calculable from such a process. 

The phenomenon I am talking about is in general known as a relaxation phenomenon. In a relaxation phenomenon, you add perturbations and see how they decay. A relaxation phenomenon is the subject of nonequilibrium statistical mechanics or hydrodynamics. The important quantities there are transport coefficients. Examples are bulk and shear viscosities, speed of sound, thermal conductivity, and so on. They measure different physical properties, but they all measure how some effect propagates. This is the characteristic of transport coefficients. One can compute all these quantities from the AdS/CFT duality, but the AdS/CFT duality is especially useful to determine the ratio of the shear viscosity $\eta$ to the entropy density $s$. This is due a {\it universality}; namely, its value does not depend on the details of gauge theories.

According to the AdS/CFT duality, this ratio is always 
\be
\frac{\eta}{s} = \frac{\hbar}{4\pi k_{\rm B}}
\label{eq:universality}
\ee
in the strong coupling limit. This result has first been shown for the ${\cal N}=4$ SYM (super-Yang-Mills theory)\cite{Policastro:2001yc,Policastro:2002se}, which is the simplest version of the AdS/CFT duality. This theory is conformal, {\it i.e.}, a theory with no scale. But the result has been extended into various theories, and the result is true for all known examples: \cite{Kovtun:2003wp}\tocite{Janik:2006ft} 
\begin{itemize}
\item Nonconformal plasmas
\item Plasmas in different spacetime dimensions
\item Plasmas at finite chemical potential
\item Plasmas with flavor
\item Time-dependent plasma
\end{itemize}

  
As one can see from \tabl{eta_over_s}, RHIC in fact suggests that QGP has a very low viscosity, which is close to the AdS/CFT value. On the other hand, the naive extrapolation of the perturbative QCD gives a much larger value. 
So, the perturbative QCD seems inaccurate and QGP seems strongly-coupled. This AdS/CFT may be useful to analyze QGP further. [See Refs.~\citen{Natsuume:2007qq,Natsuume:2007zz,Son:2007vk,Mateos:2007ay,Myers:2008fv} for further details.]

\begin{table}
\begin{center}
\begin{tabular}{ll}
AdS/CFT duality:  & $\eta/s =1/(4\pi)$\\
RHIC:\cite{Teaney:2003kp}  & $\eta/s \sim O(0.1)$\\
Naive extrapolation of the perturbative QCD:  & $\eta/s \sim O(1)$\\
Lattice result for a pure gauge theory:\cite{Nakamura:2004sy,Meyer:2007ic}  & $1<4\pi\eta/s<2$ for $1.2\,T_c<T<1.7\,T_c$ 
\end{tabular}
\caption{Various estimates of $\eta/s$.}
\label{table:eta_over_s}
\end{center}
\end{table}

\section{Brief review of causal hydrodynamics}\label{sec:causal}

\subsection{Basic idea of causal hydrodynamics}

The hydrodynamic description of QGP using the AdS/CFT duality is very successful, but this cannot be the end of the story:

\begin{itemize}

\item First, standard hydrodynamics (first-order theory) \cite{Eckart:1940te,LL} has many severe problems; {\it e.g.}, it does not satisfy causality.
\item One can restore causality, but one is forced to introduce a new set of transport coefficients. The resulting theory is known as ``causal hydrodynamics." It is also known as ``second-order hydrodynamics" or ``Israel-Stewart theory." \cite{Imuller,Israel:1976tn,Israel:1979wp} [See Refs.~\citen{Israel_review,Maartens:1996vi,Muller:1999in,Muronga:2003ta,Andersson:2006nr} for reviews.]
\item Such coefficients may become important in the early stages of QGP formation, but little is known about these coefficients. 
\end{itemize}

Causal hydrodynamics has been widely discussed in the heavy-ion literature.%
\footnote{For a recent discussion, see, {\it e.g.}, Ref.~\citen{Bhalerao:2007ek} and references therein.} 
For example, a number of groups recently reported the results of QGP simulations using causal hydrodynamics. \cite{Romatschke:2007mq}\tocite{Song:2008si} It is not my purpose to review these works, but some groups claim that $\eta/s$ must be smaller than $1/(4\pi)$ to fit the experiment. If true, this would give a serious problem to the AdS/QGP program, but the there are several potential problems. One possible problem is that they use the AdS/CFT value for $\eta/s$, but they use the weak coupling result for causal hydrodynamics. This is understandable since almost none is known about causal hydrodynamics. But is it really fine? So, we will study it using the AdS/CFT duality. 

In order to explain causal hydrodynamics, let me first remind you of standard hydrodynamics itself. Actually, hydrodynamics is rather complicated; the main object of hydrodynamics is the energy-momentum tensor, and it has two spacetime indices which complicates the story. Instead I use a charge diffusion as an example. 

There are two basic equations for the charge diffusion:
\bea
\mbox{conservation equation:}\quad && \del_\mu J^\mu=0~, 
\label{eq:conservation}\\
\mbox{``constitutive equation":}\quad && J_i= -D \del_i \rho~.
\label{eq:constitutive}
\eea
The constitutive equation tells that a charge gradient produces a current. Such a law is widely known in nature; Ohm's law is another example. For the charge diffusion, it is known as ``Fick's law." The proportionality constant is the diffusion constant $D$; it is a definition of the diffusion constant. If one combines these equations, $\rho$ and $J_i$ decouple, and one gets the equation for $\rho$ only:
\be
\del_0 \rho - D \del_i^2 \rho = 0~,
\label{eq:diffusion}
\ee
which is nothing but the diffusion equation. A propagating solution $\rho \propto e^{-iwt+iqz}$ leads to the dispersion relation 
\be
w = iD q^2~.
\label{eq:dispersion1}
\ee
This is the form we will encounter over and over again.

Hydrodynamic case is similar. Again one has the conservation equation and the constitutive equation:
\begin{subequations}
\bea
&& \del_\mu T^{\mu\nu} = 0 ~, \qquad \\
&& T_{ij} = P \delta_{ij} -\eta(\del_i u_j + \del_j u_i - \frac{2}{3}\delta_{ij}\del_k u_k) - \zeta \delta_{ij}\del_k u_k~.
\label{eq:EM_constitutive}
\eea
\end{subequations}
The first term of \eq{EM_constitutive} represents the familiar pressure part of the energy-momentum tensor, and the rest represents the dissipative part; this part defines the shear viscosity and the bulk viscosity. For the charge diffusion, $\rho$ and $J_i$ decoupled. Similarly, carry out the tensor decomposition. In this case, one gets the scalar, vector, and tensor modes. Again they decouple from each other. The scalar mode is often called as the ``sound mode," and the vector mode is called as the ``shear mode." The transport coefficients appear in the modes in \tabl{tensor_decomposition}. 

\begin{table}
\begin{center}
\begin{tabular}{|c|l|l|}
\hline
currents & tensor decomposition & transport coefficients \\
\hline
$J_\mu$
	& scalar mode (``diffusive") & $D$ \\
	& vector mode & - \\
\hline
$T_{\mu\nu}$
	& scalar mode (``sound") & $v_s$, $\eta$, $(\zeta)$ \\
	& vector mode (``shear") & $\eta$ \\ 
	& tensor mode & - \\
\hline
\end{tabular}
\caption{Tensor decomposition and the transport coefficients which appear in each channels. Here, $v_s$ and $\zeta$ are the speed of sound and the bulk viscosity, respectively. ($\zeta$ vanishes for conformal theories, so it is written in parenthesis.)
}
\label{table:tensor_decomposition}
\end{center}
\end{table}

Now, let us go back to the diffusion equation. Mathematically, \eq{diffusion} is  parabolic: it has first derivative in time but has second derivatives in space. Such an asymmetry is important for the physics of diffusion, but this means that it does not satisfy causality. For example, look at the familiar solution of the diffusion equation in $(1+1)$-dimensions:
\be
\rho \sim \frac{1}{\sqrt{4\pi Dt}}\exp\left(-\frac{x^2}{4Dt}\right)~.
\ee
This solution starts with the delta-function distribution, and it is smeared as the time passes. The point is that
it is nonzero everywhere. In particular, it is nonvanishing even outside the lightcone, so it does not satisfy causality. Another way to see the acausality is to look at propagation speeds.  From the dispersion relation (\ref{eq:dispersion1}), both the phase and group velocities diverge like $q$ with no bound. It is clear that the problem comes from the asymmetry in the number of derivatives in \eq{diffusion}. Thus, in order to restore causality, one needs a hyperbolic equation such as the Klein-Gordon equation. 

So, what is wrong? We have two equations, the conservation equation and Fick's law. The conservation equation must be true, so what is wrong is Fick's law. So, modify Fick's law as follows: 
\be
\tR \del_0 J_i + J_i= -D \del_i \rho~.
\label{eq:modified}
\ee 
The first term is a new term added, and $\tR$ is a new parameter. When $\tR=0$, it gives the original diffusion equation. What is the physical interpretation of $\tR$? Suppose that the charge gradient vanishes at some instance, $\del_i \rho=0$ for $t=0$. Then, Fick's law (\ref{eq:constitutive}) tells that the current vanishes immediately, {\it i.e.} $J_i(t)=0$ for $t\geq 0$. But this sounds unnatural; in reality, the current should decay in some finite time period. The modified law (\ref{eq:modified}) incorporates this effect. In fact, 
\eq{modified} has the solution where the current decays exponentially: $J_i(t)=J_i(0) e^{-t/\tR}$. Namely, $\tR$ is the relaxation time for the charge current.%
\footnote{
One often calls it just as the ``relaxation time" in literature, but one had better specify what is relaxing. We will see 3 relaxation times below corresponding to different currents. Even if you focus, {\it e.g.}, on the charge diffusion, one still needs a distinction since the diffusion constant $D$ itself defines the relaxation time for the charge density for a system with size $L$, {\it i.e.}, 
$\tau_{\rho} := L^2/D$. The relaxation time $\tR$ means that the current becomes {\it dynamical} in causal hydrodynamics.}
The standard diffusion equation is the case where the relaxation happens very fast so that you can approximate $\tR=0$.

Now, combine the conservation equation with the modified law: in this case, one gets the telegrapher's equation:
\be
\tR \del_0^2\rho + \del_0 \rho - D \del_i^2 \rho = 0~.
\label{eq:telegrapher}
\ee
This is a hyperbolic equation. The new term is second derivative in time, so it becomes important at early time or for rapid evolution. By now, it should be clear that what we have done is just {\it an effective theory expansion in higher orders}. Hydrodynamics is just an effective theory, so phenomenologically one has infinite number of parameters. If one wants to go beyond the standard hydrodynamics, it is natural that one needs more parameters. That is why causal hydrodynamics requires more parameters.%
\footnote{
{\it A historical anecdote:}
The modification of the diffusion equation by the telegrapher's equation (\ref{eq:telegrapher}) has a long history. It first appeared in a work of Maxwell in the context of heat transfer \cite{prehistory1}. Early works based on kinetic theory include Cattaneo, Grad, Morse and Feshbach, and Vernotte. \cite{prehistory21}\tocite{prehistory24} The problem of acausality was first noted by Cattaneo. See Sec.~VIII of Ref.~\citen{Joseph} for the detailed history.
}

A combination of $D$ and $\tR$ gives a quantity with a dimension of speed: 
\be
v_{\rm \,front}^2 :=D/\tR~. 
\label{eq:v_front}
\ee
This gives the characteristic velocity for the signal propagation. The characteristic velocity determines the rate at which discontinuities in the perturbations propagate in the fluid. So, if this speed is less than the speed of light, causality is fine. Let us rewrite the telegrapher's equation in momentum space. In the hydrodynamic limit or in the low energy limit, one gets 
\be
w = - i D q^2 - i D^2 \tR q^4 + O(q^6)~.
\label{eq:dispersion2}
\ee
The first term is the familiar dispersion relation (\ref{eq:dispersion1}) we saw earlier and the second term is the correction due to causal hydrodynamics. We will obtain such a correction from the AdS/CFT duality.

Our story is so far heuristic, but Israel carried out a systematic analysis and introduced 5 new coefficients. Among them, there are 3 relaxation times. See \tabl{tensor_decomposition2}.%
\footnote{Our notations and conventions are similar to those of Ref.~\citen{Muronga:2003ta}. In particular, one often writes our $\tpi$ as $\tPi$ in literature, but we use $\tPi$ for another relaxation time.} 
The transport coefficients appear in the diffusive, shear, and sound modes. Each mode has its own relaxation time, so there are 3 relaxation times. But little is known about these coefficients. Just like hydrodynamics, causal hydrodynamics is a framework: it does not tell the values of these parameters. So, we determine these coefficients from the AdS/CFT duality. 

\begin{table}
\begin{center}
\begin{tabular}{|c|l|l|l|}
\hline
currents & tensor decomposition & first-order theory & second-order theory\\
\hline
$J_\mu$
	& scalar mode (``diffusive") & $D$ & $\tR$ \\
	& vector mode & - & - \\
\hline
$T_{\mu\nu}$
	& scalar mode (``sound") & $v_s$, $\eta$, $(\zeta)$ & $\tpi$, $(\tPi)$\\
	& vector mode (``shear") & $\eta$ & $\tpi$ \\ 
	& tensor mode & - & - \\
\hline
\end{tabular}
\caption{Tensor decomposition and the transport coefficients including the second-order theory. The quantities in parentheses vanish for conformal theories.
}
\label{table:tensor_decomposition2}
\end{center}
\end{table}

I will not show the actual formalism since it is rather complicated, but let me describe Israel's basic procedure. First, consider a state of equilibrium. Recall the first law $Tds = d\epsilon -\mu d\rho$; it tells that the entropy density $s$ is not an independent variable, but rather $s=s(\epsilon, \rho)$. Similarly, assume that the entropy current is a function of currents even for a nonequilibrium state: $s^\mu=s^\mu(T_{\mu\nu},J_\mu)$. So, just write down the most general combinations of these currents. When $s$ is first order in currents, one gets standard hydrodynamics. If you include up to second order, you get the Israel-Stewart theory. But there is one constraint: the entropy must satisfy the second law. Thus, compute $ds$ and require that $ds$ is positive-definite. For example, $ds$ roughly includes a term like $ds \sim -J^i\partial_i\rho + \cdots$. Then, $ds$ is positive-definite if you require $J_i \propto \partial_i\rho$. This is just the constitutive equation (\ref{eq:constitutive}). Namely, you determine the generic form of constitutive equations so that the second law is satisfied.

Even though Israel's formalism is rather complicated, it is enough to consider the case of linear perturbations in order to obtain these coefficients; namely, we study plasmas near equilibrium. The real analysis is still tedious, but at the end of the day, the charge diffusion and the shear mode just take the form of the telegrapher's equation (\ref{eq:dispersion2}). That is why I used the heuristic example of charge diffusion.

\subsection{FAQ}

In this subsection, I describe some further details of causal hydrodynamics.


\subsubsection{Q: Does causal hydrodynamics completely resolve the issue of causality?}\label{sec:causality}

In order to explain causal hydrodynamics, I focused on the issue of causality. However, the issue of causality cannot be completely resolved by causal hydrodynamics since causal hydrodynamics is just an effective theory. In order to check causality, one needs a dispersion relation which is valid for all energy. But then the other higher order terms can contribute, so the issue of causality can be answered only if you sum all terms in the effective theory expansion. In other words, acausality of first-order theories simply tells that you are outside the domain of validity. (Anyway, causality should be fine in the AdS/CFT duality since general relativity respects causality.)

\subsubsection{Q: Then, how is causal hydrodynamics really useful?}

The argument of \sect{causality} may raise the question how causal hydrodynamics is really useful. Actually, standard hydrodynamics has the other difficulties such as instability and lack of relativistic covariance, and these problems can be fixed by causal hydrodynamics. 

\begin{itemize}

\item {\it Instabilities:}
In a series of papers, Hiscock and Lindblom have studied this issue extensively \cite{hiscock_lindblom2,hiscock_lindblom3,hiscock_lindblom4,hiscock_olson}. They have shown that first-order theories are unstable under small perturbations for a {\it moving} fluid \cite{hiscock_lindblom2}. Causal hydrodynamics is free from this problem \cite{hiscock_lindblom3}. From a practical point of view, the instability implies that you have no control on numerical simulations as soon as viscosity is introduced. For a numerical simulation, you are forced to consider causal hydrodynamics.

\item {\it Lack of relativistic covariance:}
Even though first-order theories look covariant, they actually spoil covariance. In order to guarantee covariance, one needs to take into account some second order terms. But then one had better consider the full second-order theory. 
This problem is clear from the discussion of instabilities. Hiscock and Lindblom found the instabilities for a moving fluid, but a moving fluid must be just a change of frame from a static fluid, so one should not find new instabilities if the theory is really covariant. This is one reason why the ``Landau-Lifshitz frame" is sometimes better than the ``Eckart frame."%
\footnote{
For a nonequilibrium state, one in general has various flows associated with different currents, so the notion of the ``fluid rest frame" is ambiguous: a different flow defines a different ``fluid rest frame." Two commonly used choices are the Eckart frame and the Landau-Lifshitz frame. However, they are not just a choice of frame in the first-order theories but they are actually different theories since first-order theories lack covariance. This is one reason why some problems of first-order theories are more severe in one ``frame." }
\end{itemize}

\subsubsection{Q: Is it appropriate to truncate at second order? }

Another common question is if it is appropriate to truncate at second order: once second order terms become important, all higher order terms can be important in general. This is a common problem to an effective theory. However, in order to really know the range of the validity of an effective theory, one needs to go to higher order terms. 
For example, if second order terms are small, then the first-order theory may be enough.
Another possibility is that third order terms are small; then, it may be fine to truncate at second order. Unfortunately, neither of these possibilities are true from the AdS/CFT analysis of gauge theory plasmas in next section.

\section{Causal hydrodynamics from AdS/CFT}\label{sec:ads_causal}

Let us go back to \tabl{tensor_decomposition2}.
In the charge diffusion, a relaxation time $\tR$ appeared as a new parameter. Another relaxation time $\tpi$ appears in the sound mode as well as in the shear mode. Also, one another relaxation time $\tPi$ can appear in general, but it vanishes for conformal theories. This gives 3 coefficients of causal hydrodynamics. Israel introduced 2 other coefficients, $\alpha_0$ and $\alpha_1$, but they vanish for black holes with no R-charge (They correspond to gauge theories at zero chemical potential), so we can ignore them. This leaves $\tpi$ and $\tR$ as interesting quantities. Determining these coefficient is our primary goal. 

What lessons should we learn? Here are some questions we would like to address:
\begin{itemize}

\item Do gauge theory plasmas really fit into the framework of the Israel-Stewart theory?

\item Is there any universality or any generic feature for the relaxation time?

\item How does the relaxation time change with the coupling?

\end{itemize}

\subsection{Do gauge theory plasmas really fit into the framework of the Israel-Stewart theory?} 

Until now, I have used the words ``causal hydrodynamics" and ``Israel-Stewart theory" interchangeably, but they are actually different notions. The Israel-Stewart theory is not a unique solution to causal hydrodynamics as described in \sect{other_formalisms}. Thus, one had better ask if gauge theory plasmas fit into the Israel-Stewart theory. 

We will consider one consistency check. 
Here are the actual dispersion relations for the shear mode and for the sound mode, respectively: 
\begin{subequations}
\bea
&& w = - i D_\eta q^2 - i D_\eta^2 \tpi q^4 + O(q^6)~,
\label{eq:dispersion_shear} 
\\
&& w = v_s q 
  - i \frac{d_s-1}{d_s} D_\eta q^2
  + \frac{1-1/d_s}{2v_s} D_\eta
  \left( 2v_s^2 \tau_\pi - \frac{d_s-1}{d_s} D_\eta
   \right) q^3 +O(q^4)~,
\label{eq:dispersion_sound}
\eea
\end{subequations}
where $D_\eta := \eta/( \epsilon + p )$, and $d_s$ is the number of {\it spatial} dimensions. The relation (\ref{eq:dispersion_sound}) is valid only for conformal theories where $\zeta = \tau_\Pi = 0$. The relation for the shear mode just takes the form of the telegrapher's equation (\ref{eq:dispersion2}). The sound mode is in general more complicated, but note that $\tpi$ appears both in the shear mode and in the sound mode. So, the relaxation time $\tpi$ can be determined both from the shear mode and from the sound mode independently. This is the consistency check we will check. So, do they agree? The answer is no.

\begin{table}
\begin{center}
\begin{tabular}{|c||c|c|}
\hline
Geometry	& shear mode & sound mode \\
\hline
&& \\
~SAdS$_4$  (M2)~ 
     & $~\displaystyle{\frac{9-(9\ln3-\sqrt{3}\pi)}{24\pi T} }~$
     & $~\displaystyle{\frac{18-(9\ln3-\sqrt{3}\pi)}{24\pi T} } \sim 0.18 \fm~$ \\
&& \\
~SAdS$_5$ (D3)~ 
     & $~\displaystyle{\frac{1-\ln 2}{2\pi T} }~$
     & $~\displaystyle{\frac{2-\ln 2}{2\pi T} } \sim 0.21 \fm~$ \\
&& \\
~SAdS$_7$ (M5)~ 
     & $~\displaystyle{\frac{18-(9\ln3+\sqrt{3}\pi)}{24\pi T} }~$
     & $~\displaystyle{\frac{36-(9\ln3+\sqrt{3}\pi)}{24\pi T} } \sim 0.27 \fm~$ \\
&& \\
\hline
\end{tabular}
\caption{The relaxation time $\tau_\pi$ computed both from the shear mode and from the sound mode. Numerical values shown correspond to $T^{-1}=1$~fm.
}
\label{table:results}
\end{center}
\end{table}

Our results are summarized in \tabl{results}. %
\footnote{
These results can be summarized in a simple form by harmonic numbers. \cite{causal_membrane}
}
As you can see, the results from the shear mode and from the sound mode do not coincide. 
This suggests that one should not take the Israel-Stewart theory too literally. The problem is that the Israel-Stewart theory is really an effective theory and one should not exceeds the validity of the effective theory. 
The dispersion relation in the shear mode (\ref{eq:dispersion_shear}) is unreliable due to the contamination from the ``third-order hydrodynamics."\cite{Baier:2007ix} In the third-order hydrodynamics, $O(q^4)$ terms can appear in the telegrapher's equation (\ref{eq:telegrapher}), which can spoil the dispersion relation (\ref{eq:dispersion2}) at $O(q^4)$. Namely, the disagreement implies that third order terms are not small in this case. Similarly, the estimate of $\tR$ in the diffusive mode is also unreliable from the same problem.
On the other hand, the sound mode is free from this problem. 
Thus, we will not discuss $\tR$ further, and we will use the sound mode results for physical interpretation of $\tpi$.

\subsection{Is there any universality or any generic feature for the relaxation time?}

The next question is if there is any universality or any generic feature for the relaxation time $\tpi$. This question is important since we cannot directly compute it for QCD at this moment. Instead, we analyze several black holes or several gauge theories to find any generic behavior. We analyze Schwarzschild-AdS black holes (SAdS) in various dimensions. For example, the ${\rm SAdS}_5$ corresponds to the D3-brane in Type IIB superstring, and it is dual to the ${\cal N}=4$ SYM. The ${\rm SAdS}_4$ (${\rm SAdS}_7$) corresponds to the M2-brane (M5-brane) in 11-dimensional supergravity, and it is dual to a 3-dimensional (7-dimensional) conformal theory.

As you can see from \tabl{results}, {\it the relaxation time is not the same among different theories}. Also, 
there is no obvious universality, but their numerical values are similar. To be more specific, get some numbers. First, recall that $\hbar c \sim 197~{\rm MeV fm}$ and 197 MeV is not far from the QCD transition temperature. So, I use 1 fm for the inverse temperature. Then, $\tpi \sim 0.2\, {\rm fm}$ for all theories we consider. %

\begin{wraptable}{r}{\halftext} 
\begin{center} 
\begin{tabular}{ccc} \hline \hline 
theory & $v_{\rm \,front}$ & $v_s$ \\ \hline 
SAdS$_4$ & 0.67 & 0.70 \\ 
SAdS$_5$ & 0.62 & 0.58 \\ 
SAdS$_7$ & 0.54 & 0.44 \\ \hline 
\end{tabular} 
\caption{Comparison of the signal propagation $v_{\rm \,front}$ and the speed of sound $v_s$.} 
\label{table:1}
\end{center} 
\end{wraptable} 

Table~\ref{table:1} compares the speed of signal propagation and the speed of sound $v_s$ in these theories. As one can see from (\ref{eq:dispersion_shear}), the combination $D_\eta$ 
plays the same role as the diffusion constant $D$, so we define the speed of signal propagation $v_{\rm \,front}$ as $v_{\rm \,front}^2 := D_\eta/\tpi$ just like \eq{v_front}. Then, $v_{\rm \,front}$ is not far from $v_s$. 


\subsection{How does the relaxation time change with the coupling?} 

The last question is how the relaxation time changes as you change the coupling constant. We use the AdS/CFT duality, so our computations are strong coupling results. There is one weak coupling estimate, so let us compare the weak coupling result with the strong coupling results.

Israel and Stewart estimated the relaxation time for a Boltzmann gas. Namely, they use the kinetic theory or the Boltzmann equation to estimate $\tpi$, so this is a dilute gas approximation. They estimated the ratio of $\tpi/\eta$ and here is the result as well as the ${\cal N}=4$ prediction: 
\bea
\frac{\tpi^{(kinetic)}}{\eta} &=& \frac{3}{2p} =\frac{6}{Ts}~, 
\label{eq:boltzmann} \\
\frac{\tpi^{({\cal N}=4)}}{\eta} &=& \frac{2(2-\ln2)}{Ts} \sim \frac{3}{Ts}~.
\label{eq:kinetic}
\eea
The first relation in \eq{boltzmann} is the Israel-Stewart result; in writing the second one, we used $\epsilon=3p$ (valid for a conformal theory) and $\epsilon+p=T s$ in order to compare it with the ${\cal N}=4$ result.
They have different functional forms, but their numerical values are not so far from each other. This suggests that the ratio $\tpi/\eta$ does not strongly depend on the coupling unlike $\eta/s$. Incidentally, it must be stressed that $\tpi$ itself does strongly depend on the coupling; this is because $\eta$ strongly depends on the coupling. But the ratio $\tpi/\eta$ does not seem to depend on the coupling strongly.%
\footnote{The finite coupling correction to $\tpi/\eta$ has been computed for the ${\cal N}=4$ theory.\cite{Buchel:2008bz} The result seems consistent with the expectation based on the Boltzmann gas approximation; the correction is positive approaching to the value (\ref{eq:boltzmann}).}

\subsection{Issue of formalism(s)} \label{sec:other_formalisms}

So far we focus on the Israel-Stewart theory since most heavy-ion literature uses the Israel-Stewart theory. But the Israel-Stewart theory may not be the most general effective theory, and there are the other candidates:
\begin{enumerate}
\item Israel-Stewart theory
\item Israel-Stewart theory modified by Ref.~\citen{Baier:2007ix}
\item ``divergence-type theories" \cite{LMR,Geroch:1990bw}
\item Carter's formalism \cite{carter}
\end{enumerate}

In fact, Ref.~\citen{Baier:2007ix} introduced 4 new coefficients $\kappa$ and $\lambda_{1,2,3}$ in addition to the coefficients which Israel-Stewart introduced. They argue that such terms are allowed in general from conformal invariance, and in some cases they are mandatory for consistency. The coefficient $\kappa$ incorporates the curved (boundary) spacetime effect, and it vanishes for gauge theories in flat spacetime. The coefficients $\lambda$'s are the coefficients in nonlinear terms; we focus on linear perturbations, and we can safely ignore them. 
It may not be surprising that the Israel-Stewart theory misses some nonlinear terms; Hiscock and Lindblom have shown that the Israel-Stewart theory fails to be hyperbolic for nonlinear perturbations \cite{hiscock_lindblom4,hiscock_olson}. 
\tabl{transport_coeffs} summarizes all transport coefficients for second-order theories known so far; there are at least 9 coefficients.

\begin{table}
\begin{center}
\begin{tabular}{|c|l|}
\hline
\multicolumn{2}{|c|}{Israel-Stewart theory} \\
\hline
$\beta_0$ (or $\tau_\Pi$), $\beta_1$ (or $\tau_J$), $\beta_2$ (or $\tau_\pi$)~ 
     & Relaxation time for viscous stress \\
$\alpha_0$, $\alpha_1$~ 
     & Coupling between $J_\mu$ and $T_{\mu\nu}$ \\
\hline
\multicolumn{2}{|c|}{Baier et al. (conformal)} \\
\hline
$\kappa$~ 
     & Effect for curved (boundary) spacetimes \\
$\lambda_1$, $\lambda_2$, $\lambda_3$~ 
     & Nonlinear terms \\
\hline
\end{tabular}
\caption{The list of transport coefficients for second-order theories. The Israel-Stewart theory has 5 coefficients, and Baier et al.\ introduced 4 more (for conformal theories)\cite{Baier:2007ix}; one needs more coefficients for nonconformal theories. 
}
\label{table:transport_coeffs}
\end{center}
\end{table}

At this moment, it is not clear how these formalisms in the list are related to each other. However, {\it if one focuses on the linear perturbations, these formalisms are all equivalent.}%
\footnote{See Ref.~\citen{Baier:2007ix} for the equivalence of formalisms~1 and 2, and Ref.~\citen{Israel_review} for formalisms~1 and 3, and Ref.~\citen{priou} for formalisms~1 and 4. The list of formalisms are far from complete. 
}
Thus,  the formalism is more or less unique for linear perturbations. So, our results based on the Israel-Stewart theory should be fine for the other formalisms as well. 
 
To summarize, we compute $\tpi$ for several theories, and we found no obvious universality: a different theory has a different $\tpi$. For practical users, here are some lessons. First, be careful when you use the Israel-Stewart theory since it is not complete (in nonlinear regime). 
 Also, $\tpi \sim 0.2\, {\rm fm}$ which is similar among the theories we consider, and the propagation speed is not far from the speed of sound. It would be interesting to see if these features are  true more in general. 
Finally, the ratio $\tpi/\eta$ does not strongly depend on the coupling, so it may be fine to keep using Boltzmann gas approximation. I hope that string theory sheds more light on this aspect of hydrodynamics since causal hydrodynamics is not completely settled. 

\section*{Acknowledgements}
I would like to thank Takashi Okamura and Tetsufumi Hirano for useful discussions. 
I would also like to thank the Yukawa Institute for Theoretical Physics at Kyoto University and the organizers of NFQCD2008 Symposium for the opportunity to present this work, and the Institute for Nuclear Theory at the University of Washington and the RIKEN-BNL Research Center at Brookhaven National Laboratory for their hospitality where this paper is partly written. This work was supported in part by the Grant-in-Aid for Scientific Research (20540285) from the Ministry of Education, Culture, Sports, Science and Technology, Japan.

%

\end{document}